\documentclass{iopjournal}

\usepackage{amsmath} 
\begin{document}

\articletype{Paper} 

\title{The Deep-Match Framework for Event-Related Potential Detection in EEG}

\author{Marek Żyliński$^{1,*}$\orcid{0000-0002-5565-0479}, Bartosz Tomasz Śmigielski$^1$\orcid{0009-0007-3336-5566} and Gerard Cybulski$^1$\orcid{0000-0002-2917-2047}}

\affil{$^1$Institute of Metrology and Biomedical Engineering, Faculty of Mechatronics, Warsaw University of Technology, Warsaw, Poland}

\affil{$^*$Author to whom any correspondence should be addressed.}

\email{marek.zylinski@pw.edu.pl}


\begin{abstract}
\textit{Objective} Reliable detection of event-related potentials (ERPs) at the single-trial level remains a major challenge due to low signal-to-noise ratio and high variability in electroencephalography (EEG) recordings. TThis work investigates the use of the Deep-Match framework for ERP detection. We examine whether incorporating prior knowledge about ERP template into deep learning models improves detection performance.
\textit{Approach} We employed the Deep-Match framework for ERP detection using multi-channel EEG recordings. The model was trained in two stages. First, an encoder–decoder architecture was trained to reconstruct input EEG signals in order to learn compact signal representations. In the second stage, the decoder was replaced with a detection module and the network was fine-tuned for ERP identification. Two model variants were evaluated: a standard model with randomly initialized filters and a Deep-MF model in which input kernels were initialized using ERP templates. Models performance was assessed on single-trial ERP detection task during leave-one-out validation.
\textit{Main results} The proposed Deep-MF model slightly outperformed the detector with standard kernel initialization for the majority of held-out subjects. Although both approaches exhibited substantial inter-subject variability, Deep-MF achieved a higher average F1-score (0.37) compared to the standard network (0.34), indicating improved robustness to cross-subject differences. Performance varied considerably across participants, with both models showing the lowest performance for one subject (F1-score = 0.01). The best performance obtained by Deep-MF reached an F1-score of 0.71, exceeding the maximum score achieved by the standard model (0.59). These results demonstrate that ERP-informed kernel initialization provides consistent improvements in single-trial ERP detection under subject-independent evaluation.
\textit{Significance} These findings demonstrate that integrating domain knowledge with deep learning architectures can substantially improve single-trial ERP detection. The proposed approach provides a step toward practical wearable EEG and passive brain–computer interface applications, enabling real-time monitoring of cognitive processes.
\end{abstract}

\section{Introduction}

An event-related potential (ERP) is a measured brain response that is evoked by a specific sensory, cognitive, or motor event \cite{luck2014introduction}. ERPs have been used in attention studies \cite{luck2000event}, for cognitive load estimation during N-back tasks \cite{brouwer2012estimating}, and for hearing threshold estimation \cite{christensen2017ear}. Changes in ERP have also been observed in schizophrenia \cite{pritchard1986cognitive}.

Since ERP amplitude is similar to spontaneously brain potentials, ERPs are burned in EEG signal. During typical ERP analysis in cognitive neuroscience, the grand average response across multiple trials is used to analyze and compare differences between ERP characteristics (amplitude, latency) across subjects \cite{CECOTTI2017156}. The requirement of several trials is mainly due to the noise in the signal coming from eye movements, muscular contractions, and ongoing brain activity that is unrelated to the experimental task. Due to averaging ERP characteristics variations across trials may not be captured by the grand average response.

On the other hand, Brain-Computer Interface (BCI) required to detect specific event-related potentials. In many real-world scenarios, repeated stimulus presentation is either impractical or impossible. For example, when stimuli appear only once—such as individual frames in a video stream or dynamically presented images—there is no opportunity to aggregate responses across repetitions \cite{cecotti2015single}. Moreover, repetition itself may affect the neural response due to memory or training effects. In such time-constrained and dynamically changing environments, decisions must rely on single-trial EEG responses. Therefore, improving the accuracy and robustness of single-trial ERP detection is essential for enabling practical, high-speed, and gaze-independent BCI systems, as well as for increasing their applicability in naturalistic settings.

The Deep Matched Filter (Deep-MF) was introduced to detect events in noisy signals. It was originally proposed for the detection of R-peaks in ear-ECG signals \cite{10416368}. Deep-MF consists of an encoder stage, initialized with an ECG template, and an R-peak classifier stage. Operating as a matched filter, the encoder searches for matches between the ECG template and the input signal. These candidate matches are subsequently processed by convolutional layers, which refine the detections and identify peaks corresponding to the ground-truth ECG. Deep-MF outperforms existing algorithms for R-peak detection in noisy ECG recordings. Moreover, it can be used for online R-peak detection on mobile devices \cite{zylinski2023hearables} and for robust heart rate estimation \cite{zylinski2025evaluate, zylinski2024hearables}.

In this paper, we evaluate the potential application of Deep-MF for single-trial ERP detection. Unlike traditional ERP analysis, single-trial detection aims to identify event-related responses within individual recordings, making the task considerably more challenging. Specifically, we assess the performance of Deep-MF in detecting pain-related ERPs evoked by laser stimulation. 

To this end we investigate whether the matched-filter-based encoder can effectively capture the characteristic temporal morphology of laser-evoked ERP components in noisy EEG recordings. To benchmark its performance, we compare the Deep-MF approach with a standard convolutional neural network (CNN) model trained directly for ERP detection.

\section{Method}

For this study we used the ds005284 dataset from OpenNeuro repository (https://openneuro.org/datasets/ds005284/). The dataset contains 64-leads EEG recorded from 26 young participants (average age 21 years, 18 Female). 

Each participant received 16 fixed-intensity laser stimuli delivered at approximately 20-second intervals \cite{lu2019music}. During the experiment, participants were instructed to keep their eyes open, focus on a fixation cross displayed on the screen, and attend carefully to each laser stimulus. After each stimulation, a 3–5 second pause was introduced, during which participants orally reported their pain rating within a 3–5 second response window. The subsequent trial began randomly within 1–3 seconds after the rating was completed.

For preprocessing of signals, we used EEGLAB. We followed established pipeline \cite{shalchy2020n}. The EEG data were filtered using high- and low-pass filters, with cut-off frequencies of 1 Hz and 40 Hz. Signals were re-referenced to the average of the P9 and P10 electrodes. Independent component analysis was then performed.

\begin{figure}[htbp]
\centering
\includegraphics[width=0.95\textwidth]{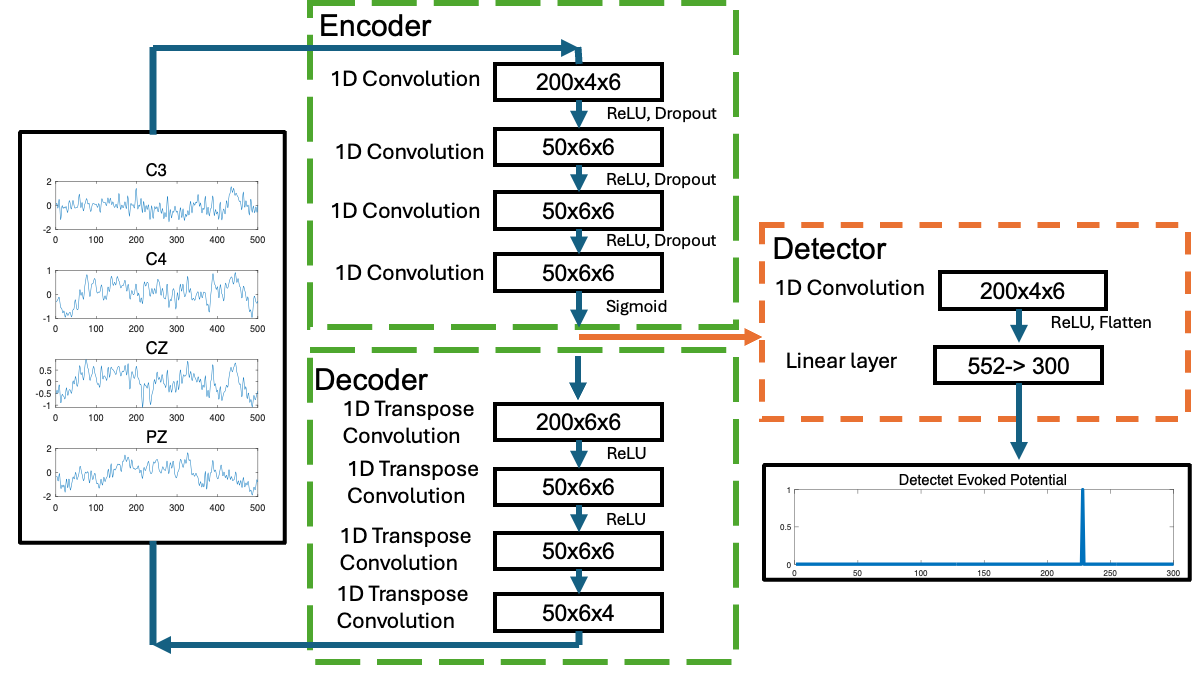}
 
\caption{Architecture of the model. The model was trained in two stages. In the first stage, the encoder–decoder network was trained to reconstruct the input signal. Two models were trained: Deep-MF, with input filters initialized using ERP templates, and a standard model with random initialization. In the second stage, the decoder part of the network was replaced with a detector, and the models were fine-tuned for ERP detection.}
\label{fig:model}
\end{figure}

\begin{figure}[htbp]
\centering
\includegraphics[width=0.8\textwidth]{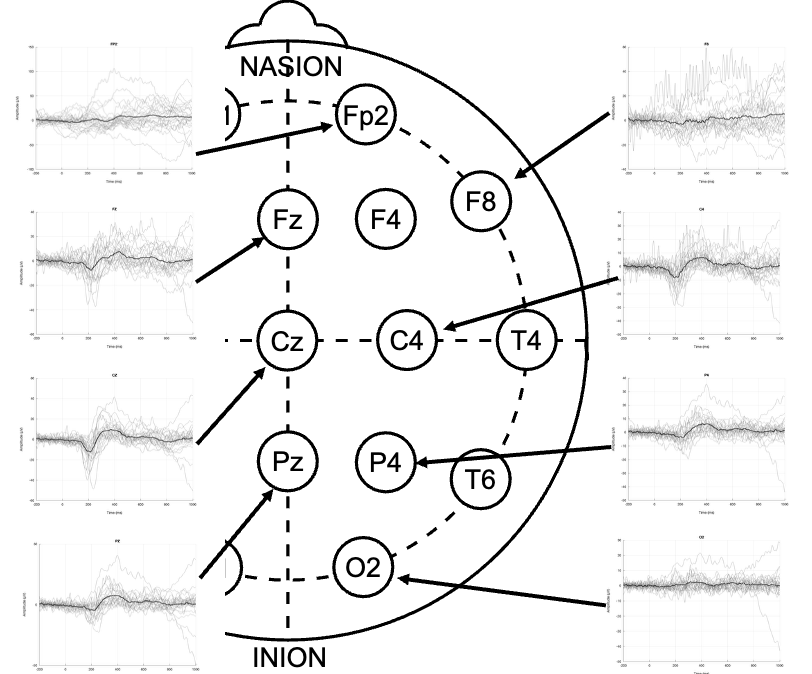}
 
\caption{Grand-average event-related potentials (ERPs) for representative EEG leads across subjects. Clear ERP components are visible over central electrodes (C3, Cz, Pz, and C4), whereas more distant electrodes show weaker or absent responses. Only a subset of leads is displayed for clarity; similar symmetrical patterns were observed across contra-lateral electrodes.}
\label{fig:potentials}
\end{figure}

First, we performed event-related potential (ERP) analysis across EEG channels. The data were epoched around the event onset, with time limits set from -0.2 s to 1 s relative to the event. The pre-event interval (-0.2 to 0 s) was used for baseline correction by subtracting the mean baseline activity from each epoch. We used the \textit{{$pop\_autorej$}} function to automatically remove epochs contaminated by artifacts. All epochs for each channel were then averaged. Figure \ref{fig:potentials} shows sample ERPs. ERP quality was spatially symmetrical, being highest at central electrodes and lower at peripheral electrodes. Based on visual inspection of the ERPs, the Cz, C3, C4, and Pz channels were selected as model inputs. For further analysis, the data were resampled to 250 Hz.

To build ERP templates for Deep-MF, the averaged ERP across all subjects was computed for each selected channel. The resulting templates were then smoothed using the \textit{smoothdata} function. In deep convolutional networks, improper weight initialization can cause trainning gradients to explode or vanish. To prevent these issues, we employed the kernel scaling method proposed by He \textit{et al.} \cite{he2015delving}. First, the mean value of each kernel was subtracted, and the templates were divided by their standard deviations. Finally, each template was scaled using the factor:
\begin{equation}
\sigma = \sqrt{\frac{2}{\text{kernel height} \times \text{kernel width} \times \text{number of input channels}}}
\end{equation}

The resulting templates were used to initialize the input kernels of DeepMF. Detectors were trained using PyTorch. All models were trained using the Adam optimizer with a learning rate of 1e-3. The architecture of the network is shown in Figure \ref{fig:model}. The model was trained in two stages. In the first stage, it was trained as an encoder–decoder \cite{occhipinti2024ear} to reconstruct the input signal. The data were standardized and split into 2-second samples with 80\% overlap. Two encoder–decoders were trained initially: one with standard weight initialization and another using DeepMF, where the input filters were initialized with ERP templates. Mean squared error loss function was used.

In the second stage, ERP detectors were trained. The pretrained models were fine-tuned, and the decoder part of each model was replaced with a detector (Figure \ref{fig:model}). The resulting models were trained to detect events, specifically the timing of the ERP. In this step, a balanced dataset was used, with an equal number of samples containing events and samples without events. For each event, 12 samples were selected with a 0.1-second step, and the starting position for each window was randomized within the step size. Samples without events were randomly selected from the dataset. As model input, 2-second segments were used. The target outputs were created as 300-sample vectors, where a value of 1 corresponded to the event time. 

In the second stage, leave-one-out (LOO) validation was used to fine-tune the encoder–decoders. For each iteration, one participant was held out as the validation set while the remaining participants were used for training. This procedure was repeated until every participant had served as the validation set once. To account for the rarity of positive events, the models were trained using a weighted binary cross-entropy loss with a positive class weight of 3.0.

For evaluation during leave-one-out (LOO) validation, ERP peaks were detected from the model outputs using a peak-finding algorithm. Detected peaks were identified with a minimum height threshold (0.25) and a minimum distance between peaks (30). The detected peaks were then compared to the ground-truth event times using a tolerance window to account for slight timing differences. True positives (TP), false positives (FP), and false negatives (FN) were computed for each participant. F1 metric was used to assess the accuracy and reliability of the ERP detection.

All scripts used for the analysis reported in this study are available in the GitHub repository: \url{https://github.com/Marower/DeepMF_ERP_Detection_in_EEG-main}.

\section{Results}

ERPs were analyzed across multiple EEG leads to examine their spatial distribution. As illustrated in Figure \ref{fig:potentials}, clear ERP components are consistently observable over the central electrodes (C3, Cz, and Pz) across subjects. In contrast, electrodes located further from the central region exhibit markedly weaker or indistinct ERP responses. This spatial pattern suggests that task-related neural activity is predominantly localized over the central scalp areas. For clarity, only a subset of EEG leads is shown in the figure; however, similar symmetrical patterns were observed across the corresponding contralateral electrodes.

Based on these observations, four channels (C3, Cz, Pz, and C4) were selected for subsequent analysis. Restricting the analysis to these channels reduces input dimensionality while preserving the most informative signals, as a result, reduce network size, improving computational efficiency.

\begin{figure}[htbp]
\centering
\includegraphics[width=0.85\textwidth]{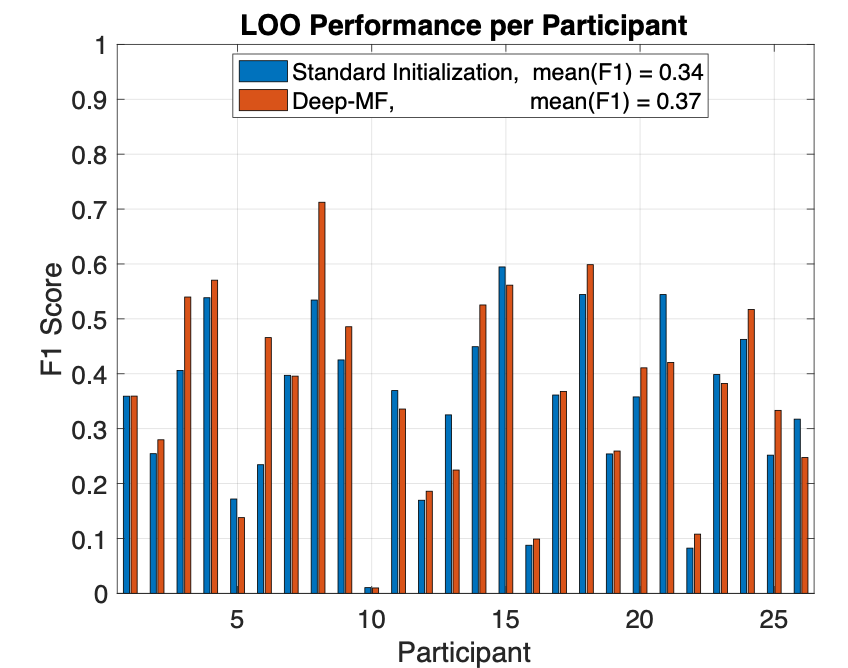}
 
\caption{F1-scores obtained for each subject using leave-one-subject-out cross-validation. Deep-MF consistently outperformed the standard network on the majority of subjects and achieved a higher mean F1-score (0.37 vs. 0.34). The lowest performance for both methods occurred for Subject 10 (F1 = 0.01).}
\label{fig:LOO}
\end{figure}

Leave-one-subject-out cross-validation was performed to evaluate subject-independent generalization. Figure \ref{fig:LOO} presents the F1-scores obtained for each held-out subject. Deep-MF slightly outperformed the detector trained with standard kernel initialization on the majority of subjects. Although both methods exhibited variability across individuals, Deep-MF achieved a higher average F1-score (0.37) compared to the second detector (0.34), suggesting improved robustness to inter-subject differences.

Both methods showed their poorest performance on Subject 10, where the F1-score was 0.01. The highest F1-score achieved by Deep-MF was 0.71, compared to 0.59 for the standard network.

\section{Discussion}

ERPs exhibit substantial variability across single trials. Differences in amplitude and latency carry important information regarding attention, mental state, fatigue, and habituation \cite{lange2005method}. Single-trial EP evaluation remains a challenging problem. Several approaches have been proposed to address this issue. ERPs can be estimated using iterative template-matching methods \cite{woody1967characterization}, adaptive filtering and weighted averaging techniques \cite{davila2002weighted}, or matched filtering approaches \cite{woodworth2007detection}. Depuydt \textit{et al.} \cite{depuydt2023single} demonstrated that neural networks outperform conventional single-trial latency estimation methods.

In our study, the Deep-MF has presented slightly better performance that standard CNN model in the detection of ERP. Task required to provide exact time of ERP, not only binary classification. However, the overall performance of the model is worst that in case of detection of R-peaks using Deep-MF \cite{10416368}. This can be explained by lower SNR of signal, amplitudes of evoked potentials are lower than the amplitude of the spontaneous EEG \cite{kidmose2013study}, and higher intersubject variability of the ERP template.

Shalchy \textit{et al.} \cite{shalchy2020n} found significant differences in behavioural and electrophysiological signatures in response to N-back stimulus type, task structure, preprocessing method, and laboratory equipment. Late latency peak (P300) is rather related to cognitive processing than to the physical attributes of stimulus \cite{nguyen2025cognitive}. Brain cognitive state can be affected by multiple factors, external like ambient sounds or internal arousal level \cite{fekri2023regulation}. Consequently, systematic investigation of preprocessing strategies to determine the optimal configuration remains an important task for future work.

Individual differences in ERPs may arise from multiple sources, including biological factors (age, brain anatomy, clinical status), cognitive traits (attention, working memory, executive control), momentary brain states (arousal, fatigue), task and stimulus properties, and technical aspects of EEG acquisition and preprocessing. These individual differences present major challenges for Deep-MF models, which are based on signal templates. One possible solution is the use of individualized or reinforcement learning–based templates that adapt the network to personal characteristics \cite{tian2025machine}. Development of the method will require interdisciplinary cooperation that recognizes the full range of human cognition, effort, and attention, with the goal of improving the ability of ERP detection models to adapt to changing cognitive states and ensure reliable performance in real-world conditions outside laboratory environments. Such research should include not only laboratory studies but also real-life experiments. \cite{bruya2018attention}.

Another approach is to improve detection by analyzing the internal structure of ERP signals and designing multiple matched-filter templates that capture their characteristic components. Longe \textit{et al.} \cite{lange2002segmented} for movement related potential the ERP can be divided into three main components: P1-N1-P2-N2 complex, P300 component and late potentials. Authors used 5 templates for matched filters and achieved SNR improved of up 10dB compared to other methods. 

Moreover, different components of an evoked potential complex may originate from different functional brain sites and can be distinguished according to their respective amplitudes \cite{lange2002modeling}. In this study ERP responses were mainly localised in central leads, but for different stimulus potentials may originate from different sites, increasing complexity of proper Deep-MF design.

The performance of deep learning methods critically depends on the quality and quantity of the available training data. This is particularly true for physiological time series, which are often noisy and limited in size, thereby motivating the use of data augmentation techniques to artificially increase dataset size \cite{moutonnet2025augmentation}. A data augmentation approach based on epoch averaging was employed by Nguyen \textit{et al.} to improve the performance of a deep learning model for single-word auditory attention decoding. Such augmentation strategies should be considered in future work to enhance the performance of Deep-MF, since the dataset used in this study contains only 16 events per subject.

The observed accuracy suggests that Deep-MF is a promising approach for reliable event detection in EEG signals. Furthermore, Deep-MF can be integrated into advanced analytical pipelines. Its computationally efficient detection capability may serve as a preliminary stage prior to the application of more complex algorithms, such as speech reception threshold evaluation \cite{borges2025speech}. By restricting the use of large-scale models, including general foundation models with millions of parameters \cite{mishra2025thought2text}, to contextually relevant signal segments, Deep-MF may reduces overall computational load. This reduction directly improves energy efficiency, which is critical for wearable systems. Furthermore, such an approach may facilitate the development of next-generation hearing assistive devices designed to reduce sustained listening effort \cite{fiedler2021hearing} through selective and attention-driven amplification.

\section{Conclusion}
In this work, we used the Deep-Match framework for the detection of event-related potentials in EEG signals. We demonstrated that the Deep-MF model outperformed the standard model. Initializing the input kernels with ERP templates improved performance and enhanced interpretability. 

Single-trial ERP detection may provide a foundation for brain–computer interface applications. Reliable single-trial ERP detection enables adaptive interfaces that dynamically adjust system behavior according to user cognitive responses. Wearable EEG headbands could detect ERP components (e.g., P300) to estimate mental workload in real time.

Future studies should focus on improving of the proposed framework. First, systematic optimization of the preprocessing pipeline may further enhance ERP detectability, as preprocessing choices strongly influence signal quality and model performance. Second, the use of adaptive ERP templates that dynamically adjust to subject-specific or session-dependent variability could improve robustness across individuals and recording conditions. Additionally, reinforcement learning approaches may allows adaptive optimization in real-time applications. Finally, use of data augmentation techniques, may increase generalization and reduce overfitting, in low-data EEG scenarios.

\roles{Marek Żyliński: Methodology, Investigation, analysis and writing – original draft.
Bartosz Tomasz Śmigielski: Writing – review \& editing.
Gerard Cybulski: Supervision and funding acquisition}

\data{For this study we used the ds005284 public dataset from OpenNeuro repository (\url{https://openneuro.org/datasets/ds005284/}). Code is available in the GitHub repository: \url{https://github.com/Marower/DeepMF_ERP_Detection_in_EEG-main}.}

\funding{Not applicable.}

\bibliographystyle{IEEEtran}
\bibliography{refs}
\end{document}